\documentstyle[12pt]{article}
\newcommand{\Pc}{Poincar\'{e} }
\newcommand{\2}{2+1}
\title{\nopagebreak
\begin{flushright}
\tenrm UCTP103.97
\end{flushright}\vskip .7in
\large \bf  Exact Local Supersymmetry, Absence of \\
Superpartners, and
Noncommutative Geometries\footnote{presented
at the Twenty Fifth Coral Gables Conference, Miami, Florida,
January 23-26, 1997}}
\author{Freydoon Mansouri\thanks{e-mail
address: mansouri@uc.edu} \\
\small \it 
Physics Department, University of
Cincinnati \\
\small \it Cincinnati, OH 45221, USA}
\date{}

\begin{document}
\maketitle

\begin{abstract}
It is pointed out that if we allow for the possibility of a
multilayered universe, it is possible to maintain exact
supersymmetry and arrange, in principle, for the vanishing of the
cosmological constant. Superpartners of a known particle will
then
be associated with other layers of such a universe. A concrete
model realizing this scenario is exhibited in \2 dimensions, and
it
is suggested that it may be realizable in $3+1$ dimensions. The
connection between this nonclassical geometry and noncommutative
geometries is discussed.
\end{abstract}

\section{Introduction}
Supersymmetry provides a rich and elegant theoretical framework
for treating fermions and bosons on the same footing. It has been
the basis of many developments for over two decades ranging from
supersymmetric quantum mechanics [1] and supersymmetric gauge
theories [2] to superstring theories [3]. The usefulness of this
concept
as an approximate symmetry in atomic [4] and nuclear physics [5]
is
already indisputable. What is not yet clear is whether it is a
symmetry of Nature at the most fundamental level, and if so in
what form.

From a purely theoretical point of view, the rich mathematical
structure of supersymmetry has been used to address a number of
important unsolved physical problems. Among these are the gauge
hierarchy problem, the cosmological constant problem, and the
dark matter
problem. Moreover, making use of such concepts as duality,
holomorphicity, etc., 
supersymmetric gauge theories can be used to analyze the dynamics
of gauge theories exactly [6]. This permits, among other things,
a
new approach to solving the longstanding strong coupling problem.
Since the dynamical mechanisms made use of in these developments
are standard to all gauge theories, it is hoped that they will
also be applicable to non-supersymmetric gauge theories. It is
thus
clear that in the absence a competing framework general enough to
address all of these problems, the optimism in the relevance of
some form of supersymmetry in a fundamental theory is not
unreasonable.

One serious drawback of supersymmetry as a fundamental symmetry
is its lack of experimental support. Up to the presently
available energies, there is no evidence for the existence of the
superpartners of the known fundamental particles such as the
electron and the photon. The standard interpretation of the
absence of superpartners is to assume that supersymmetry is
spontaneously broken and that, as a result, the superpartners
acquire large masses, making them undetectable at currently
accessible energies. It then follows that the absence of
superpartners is only temporary and that experiments at a high
enough energy scale will eventually lead to their discovery.
Depending on the particular model, a typical lower bound for such
a scale is of the order of Tev's. Unfortunately, there are no
reliable upper bounds for this scale below the Planck scale.

From the experience with flavor symmetry and the manner in which
different generations of quarks and leptons were predicted and
discovered at higher and higher energies, it is generally
believed that if supersymmetry plays a fundamental role, the 
above interpretation is the logical next step beyond the
bosonic symmetries in particle physics. On the other hand, in a
broader perspective, the consequences of the manner in which
supersymmetry is broken are not confined to the particle physics
sector. They will also have profound cosmological
consequences. In
particular, if supersymmetry is spontaneously broken in the usual
way, one would have to look for a different mechanism to ensure
the vanishing of the cosmological constant. So, if we look to
supersymmetry as basis for the simultaneous solution of both the
cosmological constant problem and the absence of superpartners,
we appear
to have reached an impasse.

A way out of this dilemma was suggested by Witten [7] based on
how local supersymmetry is realized in \2 dimensions. There is
also an alternative unconventional suggestion by my collaborators
and me [8], which was again first encountered in connection with
how local supersymmetry was
realized in \2 dimensions. In the following sections, I will
describe, in turn, these suggestions, how the alternative view
was
arrived at, its connection to noncommutative geometry, and some
of its consequences.

\section{Witten's Observation}
As mentioned in the in previous section, the jest of Witten's
observation is that in \2 dimensions the requirement of local
supersymmetry provides a mechanism that, at least in principle,
ensures the vanishing of the cosmological constant without
leading to equality of masses for the supersymmetric partners
[7].
The success of this mechanism depends crucially on the manner in
which the states of nonzero energy (mass) are realized in \2
dimensions.
To see this, we note that states of nonzero energy produce
geometries which are asymptotically conical [9]. To have Fermi-
Bose degeneracy in such a conical geometry, supersymmetry must be
realized linearly, i.e., we must have asymptotic supersymmetric
generators
(supercharges) connecting fermionic and bosonic states of a
supermultiplet. On the other hand, supercharges transform as
Lorentz spinors so that their existence depends on whether the
corresponding manifold allows the construction of spinors which
are asymptotically covariantly
constant. This cannot happen in an asymptotically conical
geometry [10]. As a result, there will be no supercharges for
constructing a linear representation of supersymmetry to which
fermionic and bosonic states of nonzero mass could belong.
Therefore, there will be no  Fermi-Bose mass degeneracy. On the
other hand, the geometry produced by the vacuum state which is a
state of zero mass is not asymptotically conical, there are no
restrictions on spinors, the vacuum remains supersymmetric, and
the cosmological constant can be made to vanish.

If it were possible to implement this mechanism in $3+1$
dimensions in a realistic manner, it would significantly boost
our
near term confidence in the relevance of supersymmetry as a
fundamental symmetry. The only obstacle on its way to full
acceptance would then be its experimental confirmation.          

\section{An Alternative Proposal}
In this section, I would like to present a point of view in which
local supersymmetry is realized as a supermultiplet of
space-times
that I will refer to as a supersymmetric space-time. The
geometry of such space-times are more complex than the familiar
classical geometries and require the introduction of new
concepts. It will be recalled that a classical metrical geometry
is determined locally in terms of a differential line element or,
equivalently, in terms of the components of a metric tensor. The
supersymmetric space-time that we have in mind is an example of a
nonclassical geometry which consists of the following elements :
(i) The c-number line element is replaced with an ``operator''
line element. Equivalently, the components of the metric tensor
are replaced with operators. (ii) These operators are constructed
from the elements of an algebra. The particular algebras of
interest
in the present context are Lie and super Lie algebras. (iii)
There is an associated Hilbert space on which the elements of the
algebra act linearly. For
a supersymmetric space-time, the corresponding Hilbert space is a
supersymmetric multiplet realizing, say, the super \Pc group.

The classical, long wave long wave length, limit of these
nonclassical geometries can be determined by allowing the line
element operator to act on the states of the associated Hilbert
space. Then the diagonal elements
may be replaced by the corresponding eigenvalues. As a result,
for each state of the Hilbert space, the line element operator
produces a ``layer'' of classical space-time, the number, n, of
layers being equal to the dimension of the (super) multiplet. The
off-diagonal elements of the line element operator provide the
means of communication among various layers. Thus in this limit,
a nonclassical geometry consisting of n layers of d-dimensional
classical geometries may be viewed as a (d+1)-dimensional
geometry
in which the range of one of the dimensions is finite and
discrete.
As an example, consider an $N=2$ supersymmetric space-time. It 
consists of four layers of d-dimensional space-times in which
different layers are related to each other by supersymmetry
transformations. We will see below a concrete realization of this
nonclassical geometry in \2 dimensions. 

Although the nonclassical geometry described above appears to be
general and independent of the dimension d, it is conceivable
that, like the mechanism suggested by Witten [7], it will only
have
\2 dimensional realizations. But for the moment let us assume
that it will also have $3+1$ dimensional realizations and
consider some of its predictions. Representing a particle by a
\Pc
state, we can put such a particle and its superpartners in a
supermultiplet consisting of these \Pc states. Then the above
nonclassical geometry, in its simplest form, suggests that the
particle and its superpartners reside in different layers of the
supersymmetric space-time. Since, in the simplest model,
supersymmetry is the only means of communication between the
layers, to have any hope of obtaining information about the
superpartners, supersymmetry must remain exact. From this it
follows that a particle and its superpartner(s), if they can be
called that, must have the same mass and that the cosmological
constant problem is, in principle, solvable.

An immediate difficulty with this picture which comes to mind is
that there is no experimental evidence for superpartners of the
same mass. In this respect, we must note that the experiments in
question were all perceived and carried out under the assumption
of a single layered Universe. So, one would not expect to obtain
any
information about the superpartners which ``reside'' in the other
layers. Moreover, the very notion of a ``superpartner'' makes
sense in a world of broken supersymmetry. In a superworld of
exact supersymmetry, a particle and its superpartner(s) are
different spin states of the same ``superparticle''. So one way
of restating the lack of experimental evidence for the mass
degenerate superpartners is to ask why it is that only one spin
state of a superparticle appears in our experiments. A possible
answer to this question is that a multilayered universe which
emerges from a nonclassical geometry is very much like the many
worlds picture necessary for an objective interpretation of
quantum mechanics [11]. A superparticle in a multilayered
universe
is
capable of being in any one of its spin states. In an experiment
set up in any one layer, the wave function of the
superparticle ``collapses'' into an eigenstate of spin consistent
with that layer. In this sense, the other spin states are
``confined.'' This makes the task of obtaining information
about superparticles highly non-trivial but not impossible. We
must learn how quantum mechanics works in such a superworld.
Needless to say, in the above discussion I have left out such
intrinsic quantum mechanical effects as tunnelling, etc. I have
also left out the possibility that for $N > 1$ there can be other
off-diagonal operators connecting the layers even when
supersymmetry is broken.

Finally, let me say a few words about a possible impact of a
supersymmetric space-time picture on the dark matter problem.
In a universe consisting of only one layer, there is
excellent experimental support for the equality of the
gravitational and inertial masses. In a multilayered universe,
there is no \'a priori reason for equivalence principle to
hold in
its present form. So, it is plausible that the need for dark
matter arises from the breakdown of the equivalence principle in
a multilayered universe. For one thing, a superparticle of a
given (inertial) mass contributes equally to the gravitational
effects in every layer of this superworld.

\section{Works on Noncommutative Geometries}
The basic element of the nonclassical geometry described in the
previous section is the introduction of an operator line element.
The simplest way of viewing such an operator is to take the
components of the metric tensor to be (noncommuting) operators.
This statement is basis dependent, however, and a transformation
to a different basis mixes the components of the metric tensor
and the coordinates. Therefore, in the transformed basis the
coordinates also become (noncommuting) operators. This means that
we can view this nonclassical geometry as a form of a
noncommutative geometry.

The subject of noncommutative geometry has appeared in
theoretical physics in number of contexts. The most comprehensive
among these is the work of Connes [12]. From a purely
physical point of view, it has appeared in the works of Witten
[13] and of 't Hooft [14]. It is also inherent in any quantum
mechanical
matrix model, or zero-brain formalism such as the work of
reference [15] and the references cited therein. To my knowledge,
no systematic study has been undertaken to see whether or not all
of these works as well as our nonclassical geometry fall within
the general formalism of Connes. The answer to this question is
likely to accelerate the progress in this field. 

\section{Lessons from \2 Dimensions}
To provide a concrete realization of the nonclassical geometry
discussed in the previous sections, we now turn to the Chern
Simons gauge theory of the super \Pc group in \2 dimensions.
It has been known for sometime that supergravity theories in \2
dimensions can be formulated as Chern Simons gauge theories of
the corresponding supergroups [16-19].
In this and the following sections, I will explore the physical
properties of the emerging space-time when 
supersymmetric matter is coupled to these theories in a super \Pc
gauge invariant manner [20]. Let me begin with the simpler
problem
to the same aim, i.e., that of coupling matter to \Pc Chern
Simons
gravity
in a \Pc gauge invariant manner. It has been shown [20] 
that the two-body problem for this theory is exactly
solvable.
One of
the important features of this approach is that the concept of
space-time is not a fundamental input but an output of the gauge
theory.

The general form of the Chern Simons action in \2
dimensions given by
\begin{equation}
I_{cs} = \int_M \gamma_{bc}A^{b}\wedge (dA^{c}+\frac{1}{3}
f^{c}_{\;de}A^{d}\wedge A^{e})
\end{equation}
where A$^a$ are components of the Lie algebra valued connection
\begin{equation}
A = A^{a} G_{a};\;\; A^{a} = A^{a}_{\mu} dx^{\mu}\end{equation}
The quantities G$^a$ are elements of the Lie algebra with
structure
constants f$_{abc}$.  The quantities $\gamma_{ab}$ are the
components of a suitable non-degenerate metric on the Lie algebra
[17].  For \Pc algebra with elements {P$_a$,J$_a$}, a=0,1,2, the
connection can be written as
\begin{equation}
A_{\mu}=e^{a}_{\mu}P_{a}+\omega^{a}_{\mu}J_{a};\;\;\;\;\mu=0,1,2
\end{equation}
where $e^{a}_{\mu}$ and $\omega^{a}_{\mu}$ are gauge fields of
the
\Pc group. The manifold $M$ in Eq. 1 is not to be identified with
the metrical space-time.

Consider next the coupling of the Chern Simons action to matter.
Any coupling via matter fields appear to break the local \Pc
gauge symmetry to its Lorentz subgroup, so that we are limited to
matter coupling via sources. The \Pc invariance of the Chern
Simons gauge theory
suggests that we introduce the notion of a particle or a
source as an irreducible representation of the \Pc group,
in the same way as we do in particle physics in 3+1
dimensions. Then, its first Casimir operator
$p^{2} = m^{2}$ determines the mass of the source, and its second
Casimir operator $W^{2} = m^{2} s^{2}$ its spin $s$. So, for
sources
of any spin, the coupling to the Chern Simons action can be
achieved in terms of the action [20]
\begin{equation}I=\int_{C}d\tau \; \eta_{ab}\;
[p^{a}\partial_{\tau}q^{b}+t^{\mu}(p^{a}e^{b}_{ \;
\mu}+j^{a}\omega^{b}_{
\;\mu})]+\lambda_{1}(p^{2}-m^{2})+\lambda_{2}(W^{2}-m^{2}s^{2})
\end{equation}
where $t^{\mu}=dx^{\mu}/d\tau$. It is clear from the action that
the quantities $p^{a}$, and
$q^{a}$ are canonically conjugate to each other and satisfy
Poisson brackets. For more than one source, we can add an action
of this type for each one of them.   
In the presence of sources, the topology of the manifold M is
modified, but the components of the field strength still vanish
outside sources.

The problem of two sources coupled to
the Chern
Simons gravity can be solved by reducing it to an equivalent
one-body problem [20].
This is done by taking full advantage of the topological features
of the
theory.  In a
topological
gauge theory all the gauge invariant observables can be expressed
in terms of Wilson loops. This means that the Casimir invariants
of a \Pc state, which we
identify as mass and spin, must be Wilson loops. Thus we can view
our gauge invariant observables of this theory as the Casimir
invariants of an 
equivalent one-body \Pc state. Such a source is source endowed
with two charges: a charge {$\Pi^{a} = 
(\Pi^{0}, \vec{\Pi})$\nolinebreak} and a charge {\nolinebreak
$\Psi^{a} = ( \Psi^{0}, \vec{\Psi})$,} such that the Casimir
invariants of the corresponding state are given, respectively, by
{\nolinebreak$\Pi\cdot \Pi = H^{2}$} and {\nolinebreak$\Pi\cdot
\Psi = HS$}. We identify $H$ and $S$ as the mass and spin of the
one-body source and wish to 
evaluate them in terms of Wilson loops of the two body system. 
For two sources with charges ($p^{a}_{1},\; j^{a}_{1}$)
and ($p^{a}_{2}, \;j^{a}_{2}$), respectively, the explicit
evaluation of the Wilson loops
were carried out in reference [20]. Here we quote the expression
for $H$ :
\begin{equation}
\cos\frac{H}{2}=\cos(\frac{m_{1}}{2})\cos(\frac{m_{2}}{2})-\frac{
p_{1}\cdot p_{2}}{m_{1}m_{2}}
\sin(\frac{m_{1}}{2})\sin(\frac{m_{2}}{2})
\end{equation}

\bigskip
\noindent{\bf The Physical Space-Time}
\bigskip

Let us now consider the structure of space-time which corresponds
to this exact solution. Up to this point, we have constructed a
Chern Simons gauge
theory coupled to sources on R$\times \Sigma$ (x-space) which as
we emphasized is metric independent and should not be identified
with space-time. On the other hand,it is clear that the
identification of quantities such as momenta and coordinates of
physically realizable particles can only be made in a metrical
space-time. So we must show how the notion of a metrical
space-time emerges from this formalism and what our gauge
invariant observables correspond to in such a space-time [21]. To
this
end, we recall that
our two sources are characterized by charges
$(p_{1}^{a},j_{1}^{a})$ and 
$(p_{2}^{a},j_{2}^{a})$ with the corresponding canonical
coordinates 
$q_{1}^{a}$ and $q_{2}^{a}$, respectively. Without loss of
generality, 
let the first source be at rest at the origin, i.\ e.\ ,
$\vec{q}_{1}=0$.
Then $\vec{q}_{2}\equiv \vec{q}$ can be viewed as a relative
coordinate. 
We parametrize $\vec{q}$ by its polar components:
$\vec{q}=(r,\phi)$.
By fixing $\vec{q}_{1}=0$, we have made a choice of gauge
which 
fixes all the \Pc gauge transformations except for the 
spatial rotations generated by $J^{0}$ and translations along
$q^{0}$. 
To fix these, consider first the transformation
\begin{equation}
\vec{q'}=\left[ \exp i\tau^{0}J_{0} \right]\vec{q}
\end{equation}
where
\begin{equation}
\tau^{0}=(1-\frac{H}{2\pi})\phi\equiv \alpha \phi=\phi'
\end{equation}
Although $H$ is a complicated function of the dynamical variables
of the two sources, for the moment let us take it to be the
numerical value of the exact Hamiltonian given by
Eq. 5. Being an element of \Pc group, this transformation
leaves the 
Casimir invariants $H$ and $S$ unchanged. But the resulting
vector, $\vec{q'}$, is 
no longer $2\pi $ periodic and satisfies the matching conditions
for the 
coordinates on a cone characterized by the deficit angle $\beta
=H$. This can be seen by noting that the
transformed coordinates $\vec{q'}$ acquire a phase under the
rotation $\phi\rightarrow\phi + 2\pi$ : 
\begin{equation}           
\vec{q'} (\phi+2\pi )=[\exp{(2\pi - H)J_{0}}]\, \vec{q'} (\phi )
\end{equation}
To completely fix the gauge, we must also fix translations along
$q^{0}$. So, consider
\begin{equation}
q'^{0}=q'^{0}(q^{0},\phi')=q^{0}-\frac{S \phi'}{2\pi \alpha}
\end{equation}
where $S$ is the numerical value of the second Casimir invariant
of the \Pc group.

It thus follows that the general
reduction of the two-body problem to an equivalent one-body
problem
always leads, in a particular gauge, to the motion of the
relative coordinate on a cone. We know from the analysis of
metrical general
relativity [9] that point sources generate conical space-times.
For a single source, the deficit angle of the cone is determined
by the energy (mass), E, of the source. We must therefore
identify the
quantity $H$ with the total gravitational energy of the two body
system. It generates a cone over which the relative coordinate of
the reduced two body system moves. Despite their similarities,
this
cone should not be confused with the conical space of the test
particle approximation. As is clear from Eq. 5, the deficit
angle of our cone is determined by the Casimir invariant $H$
which is a highly non-linear function of the masses and the
momenta of the two sources. In terms of the gauge fixed
variables, the expression for the line element takes the form 
\begin{equation}
ds^{2}=dq'^{2}_{0}-dr^{2}-r^{2}d\phi'^{2}
\end{equation}
Or in terms of more familiar coordinates
\begin{equation}
ds^{2}=(dq^{0}-\frac{Sd\phi}{2\pi}) ^{2}
-dr^{2}-\alpha^{2}r^{2}d\phi^{2}
\end{equation}
The coordinates in these equivalent expressions are related by
Eqs. 6 and 9.
Aside from any specific significance associated with the
quantities
$H$ and $S$ in this context, (see below), Eqs. 10 and 11 are
standard expressions for the line element of a spinning cone
[9].

It is thus clear that it is not the manifold R$\times\Sigma$ (x-
space) but the q-space, M$_{q}$, from which the classical
space-time is manufactured. Once the spatial part of $q^{a}$ is
identified with the cone, relativistic invariance requires that
$q^{0}$ be identified with the "classical time". The quantity $H$
characterizing this space-time also supplies [20], as it should,
the boundary term which is necessary for the consistency of the
canonical formalism in the metrical theory [22]. Since, as we
have
noted, $H$ depends non-trivially on the momenta of the two
sources, then, because the components of the metric tensor given
by Eq.
11 also depend explicitly on the canonically conjugate variables,
i.e., $q_{1}$ and $q_{2}$, these
components will have non-vanishing Poisson Brackets with each
other.
Moreover, it follows from Eqs. 7 and 9 that in the form given
by Eq. 10 although the components of the metric are reduced to
constants, the corresponding, primed, coordinates will have
non-vanishing
Poisson brackets. This suggests that, for consistency, the
quantity $S$ in Eq. 11, which is also a boundary term, should be
replaced with $P.J/H$. This
operator acts in the Hilbert space of the one-body \Pc state.
Thus we arrive at the ``operator line element''
\begin{equation}
ds^{2}=(dq^{0}-\frac{P\cdot Jd\phi}{2\pi H}) ^{2}
-dr^{2}-\alpha^{2}r^{2}d\phi^{2}
\end{equation}
It is interesting to note that we can
still write the line element
operator in same form as that given by Eq. 10 if we define  
\begin{equation}
q'^{0}=q^{0}-\frac{P\cdot J\phi}{2\pi H}
\end{equation}
But then, as we have noted above, the coordinates in such a
generalized geometry
will no longer commute with each other.

In the classical large distance physics, the operators $H$ and
$J$ may be safely replaced with their eigenvalues. But in a short
distance quantum mechanical context, this geometrical
non-commutativity may be significant and should not be ignored.
As we will see in section 8, the operator interpretation of the
line element will turn out to
be  crucial in describing the geometry of the supersymmetric
space-time, even in the long wavelength limit.

\section{Supersources as Supersymmetry Multiplets}
This section is devoted to the description of supersources which
are to be coupled to the Chern
Simons action for the super \Pc group. It
will be recalled [20] that in the case
of \Pc gravity the sources(particles) can be viewed as
irreducible representations
of the 
\Pc group. Similarly, we take a
superparticle(supersource) to be an
irreducible representation of the super \Pc group. From this
point of view, 
a superparticle is an irreducible supermultiplet consisting of 
several \Pc states related to each other by the action of the
supersymmetry 
generators. In the
interest of explicitness, we
will consider in detail the 
$N=2$ super \Pc group. The $N=2$  super \Pc algebra in \2
dimensions can 
be written as [23]
\begin{eqnarray}
& &[ J^{a},J^{b}] =-i\epsilon^{abc}J_{c} \hspace{.35in} ;
\hspace{.31in} [ P^{a},P^{b}]=0 \nonumber  \\
& &[ J^{a},P^{b}] =-i\epsilon^{abc}P_{c} \hspace{.31in} ;
\hspace{.31in} [ P^{a},Q_{\alpha}] =0  \nonumber \\
& &[ J^{a},Q_{\alpha}] =-(\sigma^{a})_{\alpha}^{\;\beta}Q_{\beta}
\hspace{.16in} ; \hspace{.29in} [ P^{a},Q'_{\alpha}] =0  \\
& &[ J^{a},Q'_{\alpha}]
=-(\sigma^{a})_{\alpha}^{\;\beta}Q'_{\beta}
\hspace{.16in} ; \hspace{.3in} \{ Q_{\alpha},Q_{\beta}\} =0 
\nonumber \\
& &\{ Q_{\alpha},Q'_{\beta}\} =-\sigma^{a}_{\;\alpha\beta}P_{a}
\hspace{.18in} ;
\hspace{.3in} \{ Q'_{\alpha},Q'_{\beta}\} =0 \nonumber \\
& & a=0,1,2 \hspace{.96in} ; \hspace{.34in} \alpha=1,2 \nonumber
\end{eqnarray}
The indices of the 
two
component spinor charges $Q_{\alpha}$ and $Q'_{\alpha}$ are
raised and lowered by the antisymmetric metric
$\epsilon^{\alpha\beta}$ with $\epsilon^{12}=-\epsilon_{12}=1$.
the
$SO(1,2)$ matrices $\sigma^{a}$ satisfy the Clifford algebra
\begin{equation}
\{\sigma^{a},\sigma^{b}\}=\frac{1}{2}\eta^{ab}
\end{equation}
where $\eta^{ab}$ is the Minkowski metric with signature
$(+,-,-)$.
We also have
\begin{equation}
\sigma^{a}_{\;\alpha\beta}=(\sigma^{a})_{\alpha}^{\;\gamma}
\epsilon_{\gamma\beta}\end{equation}
It is convenient to take the matrices $\sigma^{a}$ to be
\begin{eqnarray}
 \sigma^{0}=\frac{1}{2} \left( \begin{array}{cc}
                               1 & 0  \\ 
                               0 & -1 \end{array} \right) \;\;\;
;
\;\;\;
\sigma^{1}=\frac{1}{2} \left( \begin{array}{cc}
                               0 & i   \\ 
                               i & 0 \end{array} \right)  \;\;\;
;
\;\;\;
\sigma^{2}=\frac{1}{2} \left( \begin{array}{cc}
                                   0 & 1 \\ 
                                  -1 & 0 \end{array} \right)
\end{eqnarray} 
The two Casimir operators of the super \Pc group are given by
\begin{eqnarray}
C_{1}&=&P^{2}=\eta^{ab}P_{a}P_{b}  \\
C_{2}&=&\eta^{ab}P_{a}J_{b}+\epsilon^{\alpha\beta}Q'_{\alpha}
Q_{\beta}
\end{eqnarray}
The first of these is the same as the Casimir operator of the \Pc
subgroup, so that its eigenvalues can be identified with the
square of the mass of 
the superparticle. Since the Pauli-Lubanski operator (or its
square) does 
not commute with supersymmetry transformations, it must be
supplemented 
with the second term on the right hand side of Eq. 19 to
obtain a super \Pc 
invariant. We will designate its eigenvalues as $mc_{2}$.

Irreducible representations of the $N=2$ super \Pc
group in \2 dimensions can be constructed along the same lines as
those in 3+1 dimensions [24]. For massive states, 
without loss of generality we can work in a frame in
which 
the supermultiplet is at rest. Then
the non-vanishing
anti-commutators of 
the superalgebra simplify to
\begin{equation}
\{ Q_{1},Q'_{2}\} =\{ Q_{2},Q'_{1}\} =\frac{m}{2}
\end{equation}
Thus $Q_{\alpha}$ and $Q'_{\alpha}$, $\alpha =1,2$, form a
Clifford
algebra. 
We define a Clifford vacuum state , $|\Omega >$ by the
requirement
\begin{eqnarray}
Q_{\alpha}|\Omega >=0 & ; & \alpha=1,2
\end{eqnarray}
It is easy to verify that such a state exists within every
supermultiplet 
and that it is an eigenstate of $C_{1}$ and $C_{2}$:
\begin{eqnarray}  
C_{1}|\Omega > &=& m^{2} |\Omega >  \\
C_{2}|\Omega > &=& mc_{2} |\Omega >
\end{eqnarray} 
From the definition of the Clifford vacuum state in the rest
frame
of the 
superparticle, it follows that 
\begin{eqnarray}
C_{2}|\Omega > &=& P\cdot J |\Omega > \nonumber \\
               &=& ms^{0} |\Omega > \\
               &=& ms |\Omega > \nonumber
\end{eqnarray}
where we identify the eigenvalue, $s$, of the operator $s^{0}$
with the spin 
of the state $|\Omega >$. So, the Clifford vacuum state is a \Pc
state with 
mass $m$ and spin $s$: 
\begin{equation}
|\Omega >=|m,s>
\end{equation}
Consider, next, the states
\begin{eqnarray}
|\Omega_{1}> &=& Q'_{1} |\Omega >, \\
|\Omega_{2}> &=& Q'_{2} |\Omega >
\end{eqnarray}
and
\begin{equation}
|\Omega_{12}>=Q'_{1}Q'_{2} |\Omega >
\end{equation}
It is easy to verify that 
\begin{eqnarray}
s^{0}|\Omega_{1} > &=& (s-\frac{1}{2})|\Omega_{1}> \\
s^{0}|\Omega_{2} > &=& (s+\frac{1}{2})|\Omega_{1}>  \\
s^{0}|\Omega_{12}> &=& s|\Omega_{12}>
\end{eqnarray}
These three \Pc states together with the Clifford vacuum state
form an 
Irreducible supermultiplet of $N=2$ super \Pc group in \2
dimensions, which we
call a 
superparticle. Each supermultiplet is distinguished by its mass
$m$ and the eigenvalue $c_{2}=s$, where $s$ is the spin of the
Clifford vacuum 
state. The spins of the states within a
supermultiplet are fixed once the value of $c_{2}$ is specified.
For example, for $c_{2}=\frac{1}{2}$, the resulting $N=2$
supermultiplet is a vector 
multiplet consisting of a spin zero, two spin 1/2, and one spin
one \Pc states.

\section{Exact Solution of the Two-Superbody Problem}
It has been pointed out recently that the two-superbody problem
in $N=2$ Chern Simons supergravity is exactly solvable [23]. It
was in the process of giving a physical interpretation to this
solution that the departure from classical to nonclassical
geometry became unavoidable [8]. I will briefly sketch the two
superbody problem below and go over the supersymmetric space-time
which emerges from it in the next section.
As in Section 5, we begin with the general form of the Chern
Simons action in \2 dimensions 
given by Eq. 1. In the present case, the quantities $A^{B}$ are
the components of the Lie superalgebra valued connection
which for $N=2$ super \Pc algebra can be written as
\begin{equation}
A_{\mu}=e_{\mu}^{\;a}P_{a}+\omega_{\mu}^{\;a}J_{a}+\chi_{\mu}^{\;
\alpha}Q_{\alpha}+\xi_{\mu}^{\;\alpha}Q'_{\alpha}
\end{equation}
Then the covariant derivative is
\begin{equation}
D_{\mu}=\partial_{\mu}+iA_{\mu}
\end{equation}
Then, just as in \Pc gravity, the Chern Simons action for the
super \Pc
group can be written as
\begin{eqnarray}
I_{cs} &=& \frac{1}{2}\int_{M}\{ \eta_{bc}[e^{b}\wedge
(2d\omega^{c}+\epsilon^{c}_{\;da}\omega^{d}\wedge\omega^{a})]
\nonumber \\
& &
-\epsilon_{\alpha\beta}[\chi^{\alpha}\wedge(d-i\sigma_{a}\omega^{
a})\psi^{\beta}+\psi^{\alpha}\wedge(d-i\sigma_{a}\omega^{a})\chi^
{\beta}]\}
\end{eqnarray}
As in the case of \Pc gravity, the manifold M is
specified by its topology 
and is not to be identified with space-time which will emerge
(see below) as an output of this gauge theory.

To couple (super)sources to this Chern Simons theory, we proceed
in
a manner similar to the way sources were coupled to the \Pc Chern
Simons theory. From the discussion of the supermultiplets given
in section 6, we conclude that the
logical candidates for our supersources are the irreducible
representations 
of the $N=2$ super \Pc group. Then each supersource can be
coupled to 
the $N=2$ Chern Simons supergravity by an action of the form
[23]
\begin{eqnarray}
I_{s}=\int_{C}d\tau\{
p_{a}\partial_{\tau}q^{a}-\epsilon_{\alpha\beta}p^{\alpha}
\partial_{\tau}q^{\beta}-t^{\mu}(e_{\mu}^{a}p_{a}+\omega_{\mu}^{a
}j_{a}
-i\epsilon_{\alpha\beta}\chi_{\mu}^{\alpha}p^{\beta} \nonumber \\
+(\sigma \cdot
p)_{\alpha\beta}\xi_{\mu}^{\alpha}q^{\beta})+\lambda_{1}(p^{2}-m^
{2})
+\lambda_{2}(c_{2}-s)\}
\end{eqnarray}
where $\tau$ is an invariant parameter along the trajectory $C$.
Also, $mc_{2}$ is an eigenvalue of the second Casimir operator of
the super \Pc group, and $s$ is the spin of the Clifford vacuum
state of the supermultiplet. The
choice of the constraint multiplying $\lambda_{2}$ is crucial in
relating the eigenvalue of the second
Casimir invariant,
$c_{2}$, of the superalgebra to the spin content of a
supermultiplet. For more than one source, one can add an
action
of this type for each source. In the presence of supersources 
the topology of the manifold is modified. 
But the field strengths still vanish outside supersources,
and the theory is locally trivial.

It was shown in reference [23] that the exact gauge
invariant observables of the two-superbody system can be obtained
in terms of Wilson loops. They may be viewed as the Casimir
invariants of an equivalent one-superbody state, similar to the
equivalent one-body state of 
Chern Simons gravity. We will refer to these
invariants as $H$ and $C_{2}$. As we have seen above, their
eigenvalues determine mass(energy) and spin(angular momentum)
content the supermultiplet. They constitute
the asymptotic observables of the two-superbody system and were
given in references [23]. Here we note that the expression for
the
invariant $H$ is identical to the corresponding invariant for its
\Pc subgroup given by Eq. 5.

\section{The Physical Space-Time}
Having discussed the gauge invariant observables of the exact
two-superbody system, we now turn to the structure of the
corresponding space-time.
We take our clue from the space-time structure which emerged in
section 5 from the 
dynamics of the two-body system in \Pc Chern Simons gravity.
In the supersymmetric case, the situation 
is somewhat more complicated. To see why, we note that
in both cases we can associate our gauge invariant observables to
a reduced 
one-(super)body state. In the pure gravity case, such a state is
a single \Pc state. In the supersymmetric case it is a
supermultiplet consisting of 
several (four for $N=2$) \Pc states. As we saw in section 5, in
the case of \Pc Chern Simons theory, 
the structure of the emerging space-time and its asymptotic 
observables are completely determined by
the two (gauge invariant)  Casimir invariants of the reduced
one-body \Pc state. To see how this 
picture generalizes for the two-superbody system, we recall that
our two supersources are characterized by charges
$(p_{1}^{A},j_{1}^{A})$ and 
$(p_{2}^{A},j_{2}^{A})$ with the corresponding canonical
coordinates 
$q_{1}^{A}$ and $q_{2}^{A}$, respectively. Without loss of
generality, 
let the first supersource be at rest at the origin, i.\ e.\ ,
$\vec{q}_{1}=0$.
Then $\vec{q}_{2}\equiv \vec{q}$ can be viewed as a relative
coordinate. 
As in pure gravity, we parametrize $\vec{q}$ by its polar
components:
$\vec{q}=(r,\phi)$.
By fixing $\vec{q}_{1}=0$, we have again made a choice of gauge
which 
fixes all the $N=2$ super \Pc gauge transformations except for
the 
rotations generated by $J^{0}$ and translations along
$q^{0}$. 
To fix these, consider first the same transformation as that
specified by Eqs. 8 and 9. 
Being an element of $N=2$ super \Pc group, this transformation
leaves the 
Casimir invariants $H$ and $C_{2}$ unchanged. But again the
$\vec{q'}$ is no longer $2\pi$ periodic and satisfies the
matching conditions for the coordinates on a cone characterized
by the deficit angle $\beta=H$.

Up to this point, everything looks the same as in \Pc gravity
discussed in Section 5. 
However, differences begin to appear when we try to gauge fix
the translations along $q^{0}$. It will be recalled from our
discussion of supersources that an $N=2$ supermultiplet at rest
with Casimir invariants $H$ and $C_{2}$ consists of 
four \Pc states. Writing the eigenvalues of $C_{2}$ as $Hc$ for
the Clifford vacuum, these four states will have the following
spin eigenvalues :
\begin{eqnarray}
P\cdot J|H,c,s_{1}> &=& H(c-\frac{1}{2})|H,c_{2},s_{1}>
\\ 
P\cdot J|H,c,s_{2}> &=& Hc|H,c,s_{2}>  \\ 
P\cdot J|H,c,s_{3}> &=& Hc|H,c,s_{3}>  \\ 
P\cdot J|H,c,s_{4}> &=& H(c+\frac{1}{2})|H,c,s_{4}>
\end{eqnarray}
In the case of \Pc Chern Simons gravity, it was possible to
also fix the
gauge in $q^{0}$ direction by the transformation given by Eq. 9
which involved the spin of the \Pc state. Clearly, this is no
longer possible for a supermultiplet consisting of \Pc states
of different spin. This makes
it impossible for a single c-number line element of the form
given by Eqs. 10 and 11 to
describe all
the spin states of our equivalent one-superbody multiplet even in
the case of classical large distance gravity.
So, to describe all the spin states corresponding
to our gauge invariant observables $H$ and $C_{2}$, we must
generalize the usual notion of a c-number line element to the
``operator line element'' given by Eq. 12, which now acts on the
\Pc states making up the supermultiplet.
When this operator line
element acts on a state of a supermultiplet, we can replace, at
least for large distance physics, the operator $P.J/H$
with the spin eigenvalue of that state and hence specify the
corresponding c-number space-time. It therefore follows that the
description of all the spin
states of the equivalent one-body supermultiplet requires a
multiplet of space-times equal in number to the dimension of the
supermultiplet (four for $N=2$). With $k=1,..,4$, the line
elements for the members
of this space-time multiplet are given by
\begin{equation}
ds_{k}^{2}=(dq^{0}-\frac{s_{k}d\phi}{2\pi}) ^{2}
-dr^{2}-\alpha^{2}r^{2}d\phi^{2}
\end{equation}

The line element operator in Eq. 12 is not invariant under
supersymmetry
transformations, and it transforms in the same way as the \Pc
states within a supermultiplet. In other words, for $k=1,..,4$
the
line elements in Eq. 40 form an irreducible representation of
the
$N=2$
supersymmetry and are completely determined by the asymptotic
observables $H$ and $C_{2}$. Thus, the metrical description of
the two-super-
body system coupled to the super \Pc Chern Simons action requires
not just one but a supermultiplet of space-times. The
supersymmetry generators act as ladder operators relating
different layers of this nonclassical geometry. In its
simplest form such as in the
classical large distance regime,
this supersymmetric space-time may be viewed as an ordinary
space-time with an additional finite
discrete dimension. 

We have thus verified that the supersymmetric space-time which
emerges from the exact solution of the two-superbody in \2
dimensions is, in all details, a realization of the nonclassical
geometry discussed in section 3.

\section{Concluding Remarks}
Like Witten's suggestion described in section 2, it might be
thought that the interesting applications of the nonclassical
geometry described in this work are confined to \2 dimensions.
This may well turn out to be the case. However, it is not
difficult to conceive of $3+1$ dimensional realizations of this
geometry, which may or may not be interesting. For example,
noting the correspondence between point-like sources in \2
dimensions and infinite line sources in $3+1$ dimensions, one can
extend the supersymmetric space-time discussed in the previous
section to $3+1$ dimensions by simply adding a $dz^{2}$ term to
each
line element in Eq. 40. Work is in progress to see how one can
detect experimentally, and possibly rule out, the multilayer
effects of a supersymmetric
space-time. For one thing, this may also be a way of testing the
many
worlds picture of quantum mechanics. Much remains to be
clarified.

\bigskip
This work was supported in part by the Department of Energy under
the contract No.\ DOE-FG02-84ER40153.   

\vspace{1in}
\noindent{\bf References}

\begin{enumerate}
\item E. Witten, {\it Nucl. Phys.} {\bf B188} (1981) 513.
\item J. Wess and B. Zumino, {\it Nucl. Phys.} {\bf B70} (1974)
39.
\item M.B. Green, J.H. Schwarz, E. Witten, {\it Superstrings},
{\bf
Vol. I, Vol. II}, Cambridge University Press, 1985.
\item V.A. Kostelecky and M.M. Nieto, {\it Phys. Rev.} {\bf A32}
(1985) 1293.
\item F. Iyacello, {\it Phys. Rev. Lett.} {\bf 44} (1980) 772.
\item N. Seiberg and E. Witten, {\it Nucl. Phys.} {\bf B426}
(1994)
19, {\bf 431} (1994) 484.
\item E. Witten, {\it Int. Jour. Mod. Phys.} {\bf A10} (1995)
1247.
\item A preliminary version of our results were first reported at
{\it The
International Conference on Seventy Years of Quantum Mechanics
and Modern Trends in Theoretical Physics}, Calcutta, India 1/29-
2/2/1996, to appear in the
Proceedings, ed. P. Bandyopadhyay; Sunme Kim and Freydoon
Mansouri,
e-print gr-qc/9609037, {\it Phys. Lett.} {\bf B397} (1997) 81; F.
Ardalan,
S. Kim, F. Mansouri, {\it Int. Jour. Mod. Phys.} {\bf A12} (1997)
1183
\item S.Deser, R.Jackiw, G. 't Hooft, {\it Ann. of  Phys. }(N.Y.)
{\bf 152} (1984) 220; 
S. Giddings, J. Abbott and K. Kuchar, {\it Gen. Rel. Grav.}
{\bf 16}(1984) 751;
J.R. Gott and M. Alpert, {\it Gen. Rel. Grav.} {\bf 16}
(1984) 751.

\item M. Henneaux, {\it Phys. Rev.} {\bf D29} (1984) 2766
\item H. Everett, III, in {\it The Many Worlds Interpretation of
Quantum Mechanics}, ed. by B.S. DeWitt and N. Graham, Princeton
University Press, 1973.
\item A. Connes, {\it Non-commutative Geometry}, Academic Press,
1994
\item E. Witten, {\it Nucl. Phys.} {\bf B460} (1995) 335.
\item G. 't Hooft, Utrecht preprint THU-96/02, e-print 
gr-qc/9601014
\item T. Banks, W. Fischler, S.H. Shenker, L. Susskind, e-print
hep-th/9610043.
\item A. Achucarro and P.K. Townsend, {\it Phys. Lett} {\bf B180}
(1986) 89
\item E. Witten, {\it Nucl. Phys.} {\bf B311} (1988) 46 and {\bf
B323} (1989) 113
\item K. Koehler, F. Mansouri, C. Vaz, L. Witten, {\it Mod. Phys.
Lett.} {\bf A5}(1990) 935
\item K. Koehler, F. Mansouri, C. Vaz, L. Witten, {\it Nucl.
Phys.} {\bf B341} (1990) 167 and{\bf B348} (1990) 373
\item F. Mansouri and M.K. Falbo-Kenkel, {\it Mod. Phys. Lett.};
{\it Jour. Math. Phys.} {\bf A8} (1993) 2503; F. Mansouri, {\it
Comm. Theo. Phys.} {\bf4} (1995) 191.
\item F. Mansouri in {\it Proceedings of Twenty Third Coral
Gables
Conference}, ed. B. Kursunoglu, S. Mintz, A. Perlmutter, Plenum
Press, 1995
\item T. Regge and C. Teitelboim, {\it Ann. Phys. } {\bf 88}
(1974)
236
\item Sunme Kim and F. Mansouri, {\it Phys. Lett.} {\bf B
372}(1996) 72 
\item A. Salam and J. Strathdee, {\it Forts. Phys.} {26} (1978)
57; P.G.O. Freund, {\it Introduction to Supersymmetry}, Cambridge
University Press, 1986; J. Wess and J. Bagger, {\it Supersymmetry
and Supergravity}, second edition, Princeton University press,
1992

\end{enumerate}

\end{document}